\documentclass[11pt]{article}
\pdfoutput=1 
\usepackage{jheppub} 
\usepackage[T1]{fontenc} 

\usepackage{amsmath,amssymb,amsfonts}
\usepackage{mathtools}
\usepackage{physics}
\usepackage{tabularx}
\usepackage{bm}

\def\be{\begin{equation}\begin{aligned}}
\def\ee{\end{aligned}\end{equation}}
\newcommand{\vect}[1]{{\boldsymbol{#1}}}

\newcommand{\equref}[1]{eq.~(#1)}%
\newcommand{\secref}[1]{section~#1}%
\newcommand{\Btb}{\textsf{Btb }{}}%
\newcommand{\BtB}{\textsf{BtB }{}}%
\newcommand{\kkk}[1]{\textsf{KKK}\left(#1\right)}%
\newcommand{\kkj}[1]{\textsf{KKJ}\left(#1\right)}%
\newcommand{\kjj}[1]{\textsf{KJJ}\left(#1\right)}%
\newcommand{\kkkk}[1]{\textsf{KKKK}\left(#1\right)}%
\newcommand{\kkkj}[1]{\textsf{KKKJ}\left(#1\right)}%

\title{ Towards the higher point holographic momentum space amplitudes}

\author[a,b]{Soner Albayrak}
\author[a,c]{and Savan Kharel}

\affiliation[a]{Department of Physics, Yale University,\\ New Haven, CT 06511, USA}
\affiliation[b]{Walter Burke Institute for Theoretical Physics, Caltech,\\ Pasadena, CA 91125, USA}
\affiliation[c]{Department of Physics, Williams College,\\ Williamstown, MA 01267, USA}

\emailAdd{soner.albayrak@yale.edu}
\emailAdd{savan.kharel@williams.edu}

\abstract{
In this paper, we calculate higher point tree level vector amplitudes propagating in AdS$_4$, or equivalently the dual boundary current correlators. We use bulk perturbation theory to compute tree level Witten diagrams. We  show that when these amplitudes are written in momentum space, they reduce to relatively simple expressions. We explicitly compute four and five point correlators and also sketch a general strategy to compute the full six-point correlators.
}

\begin{document} 
\maketitle
\flushbottom

\section{Introduction}
After twenty years of its formulation, the  AdS/CFT duality remains the most concrete example of a theory of quantum gravity \cite{Maldacena:1997re, Witten:1998qj}. The correspondence has elucidated different features of quantum gravity and has helped us understand dynamics of strongly coupled quantum field theories.

The AdS/CFT duality is a correspondence between quantum theory of gravity in Anti-de Sitter spacetime and Conformal Field Theory (CFT) that lives in the boundary of such spaces. The AdS/CFT dictionary implies that each field, say $\phi_i$, in the gravitational theory corresponds to a local operator $\mathcal{O}$ in the CFT. Similarly, the spin of the bulk field is the same as the spin of the CFT operator, and the mass of the bulk field determines the scaling dimension of the CFT operator. The correspondence has been effective to study non-equilibrium phenomena in strongly coupled systems including condensed matter systems.\footnote{For a more recent introduction to AdS/CFT, see \cite{Penedones:2016voo}. For a complementary review on Conformal Field Theory see \cite{Poland:2018epd}.}

It is worth comparing the similarities and differences between Anti-de Sitter space correlators and the well-known flat space (Minkowski) amplitudes. The amplitudes in the Anti-de Sitter space are not defined in the usual fashion as in the case of flat space. Unlike flat space,  AdS space does not admit asymptotic states that are needed for the standard definition of S-matrix. However, AdS space does bear close resemblance to flat space S-matrix: Creation and annihilation operators in AdS space can be defined by changing the boundary conditions in the conformal boundary. The resultant scattering amplitudes in AdS space are then related to CFT correlation functions. Therefore, with the AdS/CFT correspondence, correlation functions can be computed by perturbative calculations that can be diagrammatically represented by Witten diagrams. The prescription for computing these correlators is very straightforward. However, after about two decades of AdS/CFT, despite the directness of the procedure, it remains a challenge to do concrete computations of correlation functions in the boundary for correlators of four and higher points.\footnote{\mbox{Early efforts in tackling this problem when the correspondence was at its infancy  can be seen here in \cite{Freedman:1998bj, Liu:1998ty, Freedman:1998tz, DHoker:1999mqo,DHoker:1999kzh}.}}

In the last few years, there has been a renewed attention in computing AdS/CFT correlators.  Many have argued that \emph{Mellin space} can remarkably simplify correlation functions in conformal field theories that has weakly coupled bulk duals \cite{ Penedones:2010ue, Paulos:2011ie,  Mack:2009gy, Fitzpatrick:2011ia,
	Kharel:2013mka, Fitzpatrick:2011hu, Fitzpatrick:2011dm, Costa:2014kfa}. While Mellin spaces are interesting, it is not easy to generalize the interesting results to higher spin fields. In this paper, we will study correlators in momentum space as a complementary approach.  Even though many exciting directions has been explored in momentum space \cite{Raju:2010by, Raju:2012zs, Raju:2012zr, Raju:2011mp, Arkani-Hamed:2015bza, Bzowski:2013sza, Bzowski:2015pba, Bzowski:2015yxv, Maldacena:2002vr,Maldacena:2011nz}, momentum space higher point correlators have not been fully explored. By computing higher point correlators, we hope to provide theoretical data that may shed light to the structures of these correlators in general.  

In a concurrent development, there has been great progress in our understanding of scattering amplitudes of gauge theories and gravity. 
 Despite the complications due to the proliferation of Feynman diagrams, the scattering amplitudes of multi-gluon processes for gauge theories and gravity exhibit remarkably simple expressions. For instance, many amplitudes whose computation once seemed untenable show remarkable simplicity and sometime can be expressed in one line \cite{Britto:2005fq, Witten:2003nn}. Several new formalisms, computation tools, and insights have been developed in the last ten years to understand this simplicity; for instance, appropriate choice of physical basis such as twistors, recursion relations, and geometric interpretation of amplitudes in term of a volume of an object such as the amplitudhedron has given fresh and deep insights about locality and unitarity \cite{ArkaniHamed:2008gz,ArkaniHamed:2009dn, ArkaniHamed:2009vw}. A natural question to ask is if one can replicate the success of scattering amplitudes in flat space of gauge theories and gravity to the study of AdS/CFT correlators.\footnote{For a comprehensive review, see \cite{Elvang:2013cua}.} In this spirit, we will focus our attention on gauge theory AdS amplitudes that are dual to the conserved current correlators in the boundary.

Here is the brief outline of the paper. We begin with a quick review of the formalism for momentum space correlators in \secref{\ref{sec:preliminaries}} where we will introduce bulk to bulk and bulk to boundary propagators for scalar and vector fields as solutions to their respective equations of motions. We will also review three point Witten diagrams, construct the ingredients essential to compute higher order functions, and move on to \secref{\ref{sec: Higher order}} where we will compute all four and five point diagrams explicitly and discuss calculation of six point diagrams with an explicit example. Finally, we summarize this work and discuss possible future directions in \secref{\ref{sec:conclusion}}.

\section{Review: momentum space perturbation theory in AdS}
\label{sec:preliminaries}
The main emphasis of this section is to quickly review the perturbation theory in Anti-de Sitter space. Our building blocks are Witten diagrams, where a typical Witten diagram is a combination of three distinctive constituents:
\begin{enumerate}
	\item external lines which connect the bulk point of the AdS to the boundary,
	\item internal lines that propagate the fields,
	\item vertices where interaction can take place.
\end{enumerate}

One can find the bulk to bulk propagators (which we will denote as \BtB) by solving the Green's function. By taking one of the points to the boundary, we can obtain the external lines, i.e. the bulk-to-boundary propagators (which we will denote as \Btb) where they correspond to propagation of some field perturbation into the bulk.  Lastly, one integrates over the bulk interaction points, denoted as $z$ below.

We will first discuss the propagators and list possible solutions as equations of motion both for scalars and vectors in AdS. Then we will use them in the simplest Witten diagram: a three point amplitude.

\subsection{Equations of motion and propagators}
For simplicity, let us restrict to the scalars. Working in Poincar\'e coordinates, one can derive the position space two point functions, i.e. \BtB for scalars in AdS, by solving
\be
\left(\Box+m^2\right) \mathcal{G}(x_1, z_1, x_2, z_2) = \frac{1} {\sqrt{-g}} \delta^d(x_1-x_2) \delta(z_1-z_2) ,\label{eq:ScalarGreens}
\ee
where our conventions are such that
\be
ds^2 = \frac{1}{z^2}\left( {dz^2 +  dx^2}\right) ~, \quad x \in \mathbb{R}^d ~, \quad z\in (0, \infty)
\ee
for the mostly positive convention for the metric at the boundary. The advantage of these coordinates is the manifest Poincar\'e invariance which makes it easy to transform the position space coordinates $x_i$ to their momentum space counterparts $k_i$.

The solution to \equref{\ref{eq:ScalarGreens}}  is given in terms of hypergeometric functions \cite{Fronsdal:1974ew}; 
and, one can obtain a \Btb from this \BtB by taking one of the end points to the boundary; say, with the limit ${z_1 \rightarrow 0}$. Instead, we will focus on the corresponding momentum space expressions and restrict ourselves to the massless case for simplicity. 

The expressions in what follows are based on the results of \cite{Raju:2010by, Raju:2011mp}. We first expand the Greens function in Fourier modes
\begin{equation}
\mathcal{G}_k (z, z^\prime) = \int d^d x~\mathcal{G}  (\vect{x}, z,  \vect{x}^\prime, z')~ e^{-i \vect{k} \cdot (\vect{x}- \vect{x}')}
\end{equation}
for which \equref{\ref{eq:ScalarGreens}} now reads as
\begin{equation}
\label{eqn:10}
z^{d+1}{\partial_z} z^{1-d} {\partial_z \mathcal{G}_k (z, z^\prime) }-z^2 \vect{k}^2 \mathcal{G}_k (z, z^\prime) = i \delta(z-z^\prime)z^{d+1}~.
\end{equation}

By solving this equation, one finds that
\begin{equation}
\label{axialpropagator}
\mathcal{G}(x, z, x', z') = -i\int \frac{d^dk}{ (2 \pi)^d} {~p~dp}~\frac{e^{i \vect{k} \cdot (\vect{x}-\vect{x}')} (z)^{d/2} J_{d/2}(pz) J_{d/2}(p z') (z')^{d/2}}{ (\vect{k}^2+p^2-i \vect{\epsilon})}\;.
\end{equation}

The appearance of Bessel function can be most readily observed from the massless Klein Gordon equation in AdS$_{d+1}$ which reads as
\be
\nonumber
0= \left( \partial_z \left(\frac {1}{ z^{d+1}} \partial_z\right) -z^{1-d} \vect{k}^2\right) \Phi_k
\ee
in momentum space. The solution depends on the chosen behavior around $z=0$ and the sign of $\vect{k}^2\equiv\eta^{ij} k_i k_j$. We list them in table~\ref{table1} where we define the \emph{positive definite norm} $k$ as  $k\equiv\sqrt{|\vect{k}^2|}$.

\begin{table}
	\centering\caption{Solution to the equations of motion for scalar fields \label{table1}}
	\begin{tabularx}{0.9\textwidth}{XX} 
		\hline\hline
		\textbf{Separation} & \textbf{Solution} \\ 
		\hline \\[-.18in]
		spacelike, $\vect{k}^2>0$ & $\phi(z)= z^{d/2} K_{d/2}(k z)$   \\ 
		timelike (normalizable), $\vect{k}^2<0$ & $\phi(z)=z^{d/2}  J_{d/2} (k z)$  \\ 
		timelike (non-normalizable), $\vect{k}^2<0$  &  $\phi(z)= z^{d/2} Y_{d/2}(k z)$ \\ 
		\hline\hline
	\end{tabularx}
\end{table}

\paragraph{Vector fields:}

\label{sec:preliminaries22}
Having quickly discussed the scalars, we can turn to the main interest of the paper: gauge fields. The action for a non-abelian gauge group in AdS is written as,
\begin{equation}
\label{action}
S= -\frac{1}{4} \int_{\mathbb{R}^d} d {\vect{x}} \int_0^\infty \frac{dz}{z^{d+1}} ~ F_{\mu \nu}^a F^{\mu \nu, a} 
\end{equation}
where, following closely the treatment of \cite{Raju:2010by}, we impose axial gauge. This reduces the action into\footnote{Following \cite{Raju:2010by} we simply add the gauge fixing term $\zeta \sum_a (A_0^a)^2$ and take $\zeta\rightarrow \infty$ to freeze $A_0^a$ at zero. We also integrated by parts and ignored the boundary term $S_{Axial}^B$ which does not alter the free equations of motion.}
\begin{equation}
S_{\text{Axial Gauge}}= \int  d^d\vect{x}~dz  \left(A_i^a \partial_\mu z^{3-d} \partial_\alpha A_j^a \eta^{\mu \alpha} \eta^{ij}- z^{3-d} A_i^a \partial_p \partial_q A_j^a \eta^{i  p} \eta^{j q} \right).
\end{equation}

In momentum space, the relevant equations of motion are,
\begin{equation}
\partial_0 z^{3-d} \partial_0 A_i^a(z) - \vect{k}^ 2 z^{3-d} A_i^a(z) =0
\end{equation}
for the Fourier modes\footnote{We will be mostly suppressing the momentum argument for the Fourier mode \mbox{$A_i^a(z)\equiv A_i^a(k,z)$}.}
\begin{equation}
A_i ( x, z) = \int d^d x~A_i^a (z) e^{i \vect{k} \cdot \vect{x}}.
\end{equation}

Similar to the scalar case, we have different solutions for different cases which we list in table~\ref{table2}.\footnote{Note that only the norm of momentum enters the solutions, hence we need polarization vectors $\vect{\epsilon}$. These polarization vectors satisfy transversality $\vect{k} \cdot  \vect{\epsilon} =0$ just like the gauge fields themselves $\vect{k} \cdot \vect{A} ( k, z)=0$: These conditions directly follow from the axial gauge choice.} Lastly, the Green's function for the vector field can be written in the following way
\begin{equation}
\label{btbvector}
\mathcal{G}_{ij}(\vect{x}, z, \vect{x}', z') = \int \frac{d^dk}{ (2 \pi)^d} p~ dp  \frac{e^{i \vect{k} \cdot (\vect{x}-\vect{x}')} z^{{\frac{d-2}{2}}} J_{\frac{d-2}{2}}(pz) J_{{\frac{d-2}{2}}}(p z') (z')^{}}{ (\vect{k}^2+p^2-i \vect{\epsilon})} H_{ij}(p,\vect{k})\;,
\end{equation}
where 
\begin{equation}
H_{ij}(p,\vect{k})=-i\left(\eta_{ij}+\frac{\vect{k}_i\vect{k}_j}{p^2}\right)
\end{equation}
and we have suppressed the color dependence.

\begin{table}
	\centering\caption{{{Solution to the equations of motion for gauge fields }} \label{table2}}
	\begin{tabularx}{0.9\textwidth}{XX} 
		\hline\hline
		\textbf{Separation} & \textbf{Solution} \\ 
		\hline \\[-.18in]
		spacelike, $\vect{k}^2>0$ & $ A_i^a(z)= \vect{\epsilon}_i^a z^{{\frac{d-2}{2}}} K_{\frac{d-2}{2}}(k z)$   \\ 			
		timelike (normalizable), $\vect{k}^2<0$ & $ A_i^a(z)= \vect{\epsilon}_i ^a z^{{\frac{d-2}{2}}}  J_{\frac{d-2}{2}} ( k z)$  \\ 			
		timelike (non-normalizable), $\vect{k}^2<0$  &  $ A_i^a(z) =\vect{\epsilon}_i^a  z^{{\frac{d-2}{2}}}  Y_{\frac{d-2}{2}}(k z)$ \\[.05in] 
		\hline\hline
	\end{tabularx}
\end{table}		

\subsection{A basic Witten diagram: three point amplitude} 
\label{sec:warming1}
Three point amplitudes are very important as they correspond to three point correlation functions in the holographic CFT's, which 
contain all the dynamical data of a CFT \cite{Bzowski:2015pba}. Additionally, these amplitudes will be the building blocks for the higher point computations in the following section.

By following \cite{Raju:2012zs}, we consider a three-point amplitude with three incoming momenta, $\vect{k}_1$, $\vect{k}_2$, and $\vect{k}_3$ with $\vect{k}_1 + \vect{k}_2 + \vect{k}_{3}= 0$, and observe that the result takes a product form of some tensor structure and the so-called triple-K integral,\footnote{ Similar conformal integrals have been computed in \cite{Bzowski:2015yxv}.} a function of three magnitudes denoted as $k_1$, $k_2$, and $k_3$:
\begin{multline}
\mathcal{A}_3= V_{123}\left(\vect{k}_1,\vect{k}_2,\vect{k}_3\right) 
\int_0^\infty  \frac{d z}{z^{d+1}} z^4 \left(\sqrt{\frac{2 k_1}{\pi}}z^{{\frac{d-2}{2}}} K_{{\frac{d-2}{2}}} (k_1 z)\right) \\\times\left(\sqrt{\frac{2 k_2}{\pi}}z^{{\frac{d-2}{2}}} K_{{\frac{d-2}{2}}} (k_2 z)\right)\left(\sqrt{\frac{2 k_3}{\pi}}z^{{\frac{d-2}{2}}}  K_{{\frac{d-2}{2}}} (k_3 z)\right) \label{3pt amplitude}
\end{multline}
where $V_{ijk}$ is the color-ordered three point vertex structure.\footnote{We follow the convention of \cite{Dixon:1996wi}.} The explicit form of color-ordered three and four point structures can be written as
\begin{equation}
\begin{aligned}
V_{ijk}(\vect{k}_1, \vect{k}_2, \vect{k}_3)\coloneqq{}&{}\frac{i}{\sqrt{2}}\left(\eta_{ij}(\vect{k}_1-\vect{k}_2)_k+\eta_{jk}(\vect{k}_2-\vect{k}_3)_i+\eta_{ki}(\vect{k}_3-\vect{k}_1)_j\right)\;,\\
V^{ijkl}_c\coloneqq{}&{} i \;\eta^{ik} \eta^{jl}-\frac{i}{2}\left(\eta^{ij} \eta^{kl}+\eta^{il} \eta^{jk}\right)\;.
\end{aligned}\label{eq: v structures}
\end{equation}

Here, we use the shorthand notation $V_{1jk}\equiv\vect{\epsilon}_1^iV_{ijk}$ and so on for the polarization vector $\vect{\epsilon}_1$. One may be concerned with this shorthand notation as $V_{1jk}$ can also refer to $V_{ijk}\evaluated_{i=1}$. However, we will not have this ambiguity as we will never be interested in individual components in this paper.

The structure $V$'s above are the same antisymmetric tensors that appears in flat space, contracted with the polarization vectors. In the above expression and what is to follow, we will suppress the color factors and indices to lighten the notation and this simplification will not affect our analysis. 

Even though \equref{\ref{3pt amplitude}} is valid in any dimensions, we will specialize to $d=3$ in the rest of the paper. Apart from its physical relevance, $d=3$ is also important from computational point of view: Bessel functions $K_{1/2}$ and $J_{1/2}$ are essentially sines and cosines hence we can easily perform the otherwise complicated radial $z$ integrals. For future use, let us define
\begin{equation}
\begin{aligned}
\kkk{p,r,s,z}\coloneqq&\sqrt{\frac{8prs}{\pi^3}}z^{11/2}K_{1/2}(p z)K_{1/2}(r z)K_{1/2}(s z)\\
\kkj{p,r,s,z}\coloneqq& \sqrt{\frac{4pr}{\pi^2}}z^{11/2}K_{1/2}(p z)K_{1/2}(r z)J_{1/2}(s z)
\\
\kjj{p,r,s,z}\coloneqq&\sqrt{\frac{2p}{\pi}}z^{11/2}K_{1/2}(p z)J_{1/2}(r z)J_{1/2}(s z)
\\
\kkkk{p,r,s,t,z}\coloneqq& \sqrt{\frac{16 p r s t}{\pi^4}}z^{6}K_{1/2}(p z)K_{1/2}(r z)K_{1/2}(s z)K_{1/2}(t z)
\\ 
\kkkj{p,r,s,t,z}\coloneqq& \sqrt{\frac{8p r s}{\pi^3}}z^{6}K_{1/2}(p z)K_{1/2}(r z)K_{1/2}(s z)J_{1/2}(t z)
\end{aligned}\label{definition of kkk and kkj}
\end{equation}
where we take a product of one $z^4$ and bunch of $\left(\sqrt{\frac{2 p}{\pi}}z^{\frac{d-2}{2}}K_{\frac{d-2}{2}}(p z)\right)_{d=3}$ and\\ $\left(z^{\frac{d-2}{2}}J_{\frac{d-2}{2}}(t z)\right)_{d=3}$ for each $K$ and $J$ respectively.\footnote{Here, $z^4$ is the vertex factor and it is critical in getting rid of the divergence}  We are considering $\kkj{p,r,s,z}$ and $\kjj{p,r,s,z}$ etc., as they correspond to the transition amplitude which is obtained by replacing some of the bulk-to-boundary legs of a Witten diagram with normalizable modes \cite{Raju:2011mp}.

We can immediately integrate these products:
\begin{equation}
\begin{aligned}
\int_0^{\infty }\frac{dz}{z^4}\kkk{a,b,c,z}&=\frac{1}{a+b+c}\\
\int_0^{\infty }\frac{dz}{z^4}\kkj{a,b,c,z}&=\sqrt{\frac{2}{\pi }}\frac{ \sqrt{c}}{(a+b)^2+c^2}
\\
\int_0^{\infty }\frac{dz}{z^4}\kjj{a,b,c,z}&=\frac{4}{\pi}\frac{a \sqrt{b c}}{ \left(a^2+(b-c)^2\right) \left(a^2+(b+c)^2\right)}
\\
\int_0^{\infty }\frac{dz}{z^4}\kkkk{a,b,c,d,z}&=\frac{1}{a+b+c+d}
\\
\int_0^{\infty }\frac{dz}{z^4}\kkkj{a,b,c,d,z}&=\sqrt{\frac{2}{\pi }}\frac{ \sqrt{d}}{(a+b+c)^2+d^2}
\end{aligned}\label{integratedz}
\end{equation}	

Going back to \equref{\ref{3pt amplitude}}, we can read off the three point amplitude
\begin{equation}
\label{eqn:21}
\mathcal{A}_3=\frac{V_{123}\left(\vect{k}_1,\vect{k}_2,\vect{k}_3\right)}{k_1+k_2+k_3}
\end{equation} 
which is seen in \cite{Anninos:2014lwa,Raju:2012zs}.

\begin{figure}[tbp]
	\centering
	\includegraphics[width=.3\textwidth,origin=c]{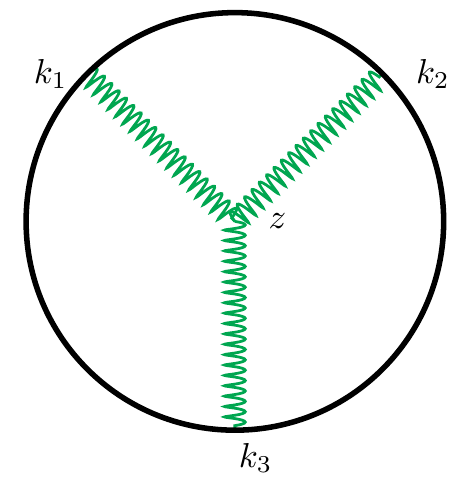}
	\caption{The three point vector amplitude. \label{3pt}}
\end{figure}

\section{Calculation of higher point amplitudes}	
\label{sec: Higher order}
As we have seen in the previous section, one can calculate any tree level diagram with the following algorithm:
\begin{enumerate}
	\item Take the product of elements of Witten diagram: $z^4$ factors for vertices, appropriately normalized\footnote{See the explanation after \equref{\ref{definition of kkk and kkj}}.} Bessel $K$ functions for the sources at the boundary, and the \BtB for internal lines. Integrate this product over the bulk point $z$'s.
	\item Using \equref{\ref{btbvector}}, decompose \BtB's over Bessel $J$ functions 
	\item Carry out the bulk point integrations: For $d=3$, we already computed the possible results in \equref{\ref{integratedz}}.\footnote{There are similar computations that are ignored as they are irrelevant unless one goes to even higher orders. For example, one needs bulk point integrations of \textsf{KJJJ} in tree level seven point diagram even though we do not consider that in \equref{\ref{integratedz}}.}
	\item Carry out the internal momenta $p$ integrations, which were introduced in the decomposition of \BtB.
\end{enumerate}

Despite the straightforwardness of the procedure, analytically carrying out the integrals of the last step can be quite cumbersome. However, we will show in this section that these integrals actually yield extremely simple results and one can use residue theorem to compute these integrations efficiently. Similarly, we will see that while the individual residues are somewhat involved the final answers are actually simple.

The validity of residue theorem lies in the fact that bulk point integrated $\textsf{KKK}$, $\textsf{KKJ}$, etc. are meromorphic functions with respect to the internal integration variable, and they vanish at infinity. Also, even though they are not individually even functions of the integration variable, the relevant integrals will always be even with respect to the internal momenta: This follows from the fact that we either have $pdp$ or $\frac{dp}{p}$ and bulk point $z$ integration of Bessel function $J$'s bring $\sqrt{p}$ at both ends of the propagator. Therefore, we can always write the integration over internal momenta using residue theorem.

In what follows, we will use the following notation to indicate norms and vectors of momenta:
\begin{equation}
k_{\underline{i_{11}i_{12}\dots i_{1n_1}}\;\underline{i_{21}i_{22}\dots i_{2n_2}}\dots \underline{i_{m1}i_{m2}\dots i_{mn_m}}j_1j_2\dots j_p}\coloneqq\sum\limits_{a=1}^{m}\abs{\sum\limits_{b=1}^{n_a}\vect{k}_{i_{ab}}}+\sum\limits_{c=1}^{p}\abs{\vect{k}_{j_c}}\;,
\end{equation}
hence, e.g. $k_{12}=\abs{\vect{k}_1}+\abs{\vect{k}_2}\;,\;k_{\underline{12}}=\abs{\vect{k}_1+\vect{k}_2}\;,\;k_{\underline{12}\;\underline{34}5}=\abs{\vect{k}_1+\vect{k}_2}+\abs{\vect{k}_3+\vect{k}_4}+\abs{\vect{k}_5}\;,\cdots$

In accordance with the above convention, we will also use
\begin{equation}
\vect{k}_{i_1i_2\dots i_n}\coloneqq\vect{k}_{i_1}+\vect{k}_{i_2}+\cdots+\vect{k}_{i_n}
\end{equation}
as a shorthand notation.

\subsection{Four point function}
\label{sec:warming2}
{In this section we will write the color-ordered four-point amplitude of a gauge boson propagating in Anti-de Sitter space at tree level}. The amplitude is the sum of three  pieces: an $s-$channel diagram, a $t-$channel diagram, and a contact diagram.\footnote{There is no $u-$channel diagram as we are considering color-ordered correlators, and color-ordered amplitudes can only have poles in channels with exchanged momenta being sum of cyclically adjacent external momenta \cite{Dixon:1996wi}.}

\subsubsection{S-channel}
\label{sec:warming21}
\begin{figure}[tbp]
	\centering
	\includegraphics[width=.3\textwidth,origin=c]{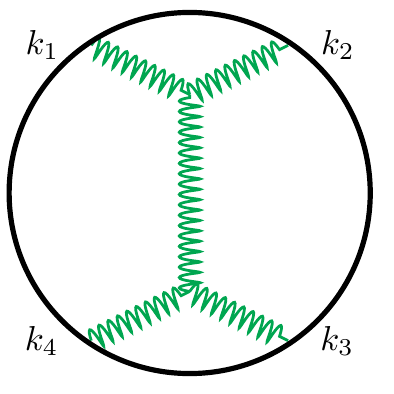}
	\caption{The four point exchange diagram.\label{4pt-schannel}}
\end{figure}

Let us start with the s-channel diagram as show in figure \ref{4pt-schannel}. The important consideration that we use while computing this integral is the factorization of the \BtB in terms of two   Bessel functions of the first kind $J_n$ as seen in \ref{btbvector}. This allows us to write down the four point function as a product of two three point function of the type $\kkj{k_i,k_j,p,z_i}$ as seen in \equref{\ref{definition of kkk and kkj}}. With these considerations,  we can write down an expression for the s-channel diagram of the four point amplitude,
\begin{equation}
\label{4pt:1}
\mathcal{M}_{4s} =\int p~dp\frac{d z}{z^{4}}\frac{d z'}{z'^{4}} \kkj{k_1,k_2,p,z}\frac{{M}^{1234}(\vect{k}_1, \vect{k}_2, \vect{k}_3, \vect{k}_4)}{ {\left( \vect{k}_{12}^2 + p^2\right)} }\kkj{k_3,k_4,p,z'}
\end{equation}
where, 
\begin{equation}
{M}^{ijkl}(\vect{k}_1, \vect{k}_2, \vect{k}_3, \vect{k}_4)= V^{ijm}(\vect{k}_1, \vect{k}_2, -\vect{k}_{12} ) H_{mn}(p,\vect{k}_{12}) V^{kln} (\vect{k}_3, \vect{k}_4, \vect{k}_{12})\;.
\end{equation}
Recall that we have $V_{12k}=\vect{\epsilon}_1^i\vect{\epsilon}_2^jV_{ijk}$ and so on in our conventions.

We would like to remind the reader that in the above expression $z$ and $z'$ are two radial coordinates where integration takes place: Note that $z^4$ and $z'^4$ are part of the appropriate volume factors in three dimensions. Writing \equref{\ref{4pt:1}} using the the bulk $z$ integral performed in  \equref{\ref{integratedz}}, we obtain
\begin{multline}
\mathcal{M}_{4s} = \int p~d p \left(\frac{\sqrt{\frac{2 p }{\pi }}}{{k_1}^2+2 {k_2} {k_1}+{k_2}^2+p^2} \right) \left( \frac{\sqrt{\frac{2 p }{\pi }}}{{k_3}^2+2 {k_4} {k_3}+{k_4}^2+p^2}  \right)\\ \times V^{12m}(\vect{k}_1, \vect{k}_2, -\vect{k}_{12} )
\frac{H_{mn}(p,\vect{k}_{12})}{\left( \vect{k}_{12}^2 + p^2\right)} V^{34n} (\vect{k}_3, \vect{k}_4, \vect{k}_{12} )\;.
\end{multline}

Let us isolate $p-$dependent parts of the above integral: We can rewrite it as
\begin{equation}
\mathcal{M}_{4s} =-i\;V^{12m}(\vect{k}_1, \vect{k}_2, -\vect{k}_{12} )  V^{34n} (\vect{k}_3, \vect{k}_4, \vect{k}_{12} ) \left(\eta_{mn}\mathcal{M}_{4s}^{(1)}+\left(\vect{k}_{12}\right)_m\left(\vect{k}_{{12}}\right)_n \mathcal{M}_{4s}^{(2)}\right)\label{decomposition of four point function}
\end{equation}
for
\begin{equation}
\begin{aligned}
\mathcal{M}_{4s}^{(1)}=& \int p~dp  \left( \frac{\sqrt{\frac{2 p }{\pi }}}{{(k_1+k_2)}^2+p^2} \right) \left( \frac{\sqrt{\frac{2 p }{\pi }}}{{(k_3+k_4)}^2+p^2}  \right) \frac{ 1}{ \vect{k}_{12}^2 + p^2 }\\ 
\mathcal{M}_{4s}^{(2)}=& \int p~dp  \left( \frac{\sqrt{\frac{2 p }{\pi }}}{{(k_1+k_2)}^2+p^2} \right) \left( \frac{\sqrt{\frac{2 p }{\pi }}}{{(k_3+k_4)}^2+p^2}  \right) \frac{ 1}{ \vect{k}_{12}^2 + p^2  }  \frac{1}{p^2 }
\end{aligned}
\end{equation}

We will compute above integrals using the method of residues. For ${M}_{4s}^{(1)}$, we have
\begin{equation}
\begin{aligned}
\operatorname*{Res}_{p=i k_{\underline{12}}} \mathcal{M}_{4s}^{(1)} &=\frac{i k_{\underline{12}}}{\pi  \left(k_{\underline{12}}^2-k_{12}^2\right)
	\left(k_{\underline{12}}^2-k_{34}^2\right)}
\\
\operatorname*{Res}_{p=i k_{12}} \mathcal{M}_{4s}^{(1)} &=\frac{i k_{12}}{\pi  \left(k_{12}^2-k_{34}^2\right)
	\left(k_{12}^2-k_{\underline{12}}^2\right)}
\\
\operatorname*{Res}_{p=i k_{34}} \mathcal{M}_{4s}^{(1)} &=\frac{i k_{34}}{\pi  \left(k_{34}^2-k_{12}^2\right)
	\left(k_{34}^2-k_{\underline{12}}^2\right)}
\end{aligned}
\end{equation}
which can be summed to a very simple expression:\footnote{Overall factor $i \pi$ follows from the residue theorem and the fact that we are integrating an even function from $0$ to $\infty$.}
\begin{equation}
\mathcal{M}_{4s}^{(1)}=i\pi\left(\operatorname*{Res}_{p=i k_{\underline{12}}}+\operatorname*{Res}_{p=i k_{12}}+\operatorname*{Res}_{p=i k_{34}}\right)\mathcal{M}_{4s}^{(1)}
=\frac{1}{k_{1234} k_{12\underline{12}} k_{34\underline{12}}}\;.
\end{equation}

Similarly, we can perform integration for ${M}_{4s}^{(2)}$. The residues are given by
\begin{equation}
\begin{aligned}
\operatorname*{Res}_{p=i k_{\underline{12}}} \mathcal{M}_{4s}^{(2)} &=-\frac{i}{\pi  k_{\underline{12}} \left(k_{\underline{12}}^2-k_{12}^2\right)
	\left(k_{\underline{12}}^2-k_{34}^2\right)}
\\
\operatorname*{Res}_{p=i k_{12}} \mathcal{M}_{4s}^{(2)} &=-\frac{i}{\pi  k_{12} \left(k_{12}^2-k_{34}^2\right)
	\left(k_{12}^2-k_{\underline{12}}^2\right)}
\\
\operatorname*{Res}_{p=i k_{34}} \mathcal{M}_{4s}^{(2)} &=-\frac{i}{\pi  k_{34} \left(k_{34}^2-k_{12}^2\right)
	\left(k_{34}^2-k_{\underline{12}}^2\right)}
\end{aligned}
\end{equation}
Again, summing over the expressions, we get:
\begin{equation}
\mathcal{M}_{4s}^{(2)}=i\pi\left(\operatorname*{Res}_{p=i k_{\underline{12}}}+\operatorname*{Res}_{p=i k_{12}}+\operatorname*{Res}_{p=i k_{34}}\right)\mathcal{M}_{4s}^{(2)}
=\frac{k_{1234 \underline{12}}}{k_{12} k_{34} k_{1234} k_{\underline{12}} k_{12 \underline{12}} k_{34 \underline{12}}}\;.
\end{equation}

\equref{\ref{decomposition of four point function}} now reads as
\begin{equation}
\mathcal{M}_{4s} =-i\frac{ V^{12m}(\vect{k}_{1}, \vect{k}_{2},-\vect{k}_{{12}})  V^{34n} (\vect{k}_{3}, \vect{k}_{4}, \vect{k}_{{12}} ) }{k_{1234} k_{12\underline{12}} k_{34\underline{12}}}\left(\eta_{mn}+\frac{k_{1234 \underline{12}}\left(\vect{k}_{12}\right)_m\left(\vect{k}_{12}\right)_n}{k_{12} k_{34}  k_{\underline{12}}}\right)\label{eq: 4s contribution}\;.
\end{equation}

\subsubsection{Contact diagram}	\label{sec:warming22}
The four point contact diagram has the same integral form as the three point amplitude in \equref{\ref{3pt amplitude}} upto the replacement of $\textsf{KKK}$ with $\textsf{KKKK}$:
\begin{equation}
\mathcal{M}_c=V^{1234}_c\int \frac{d z}{z^4}\kkkk{k_1,k_2,k_3,k_4,z}\;,
\end{equation}
where  $V^{1234}_c$ is defined in \equref{\ref{eq: v structures}}.\footnote{Here, we would like to remind the reader that $1,2,\dots$ do not refer to components in $V_c^{1234}$.} From \equref{\ref{integratedz}} we can immediately read the result
\begin{equation}
\mathcal{M}_c=\frac{V^{1234}_c}{k_{1234}}\;.\label{eq: 4c contribution}
\end{equation}		
\begin{figure}[tbp]
	\centering
	\includegraphics[width=.3\textwidth,origin=c]{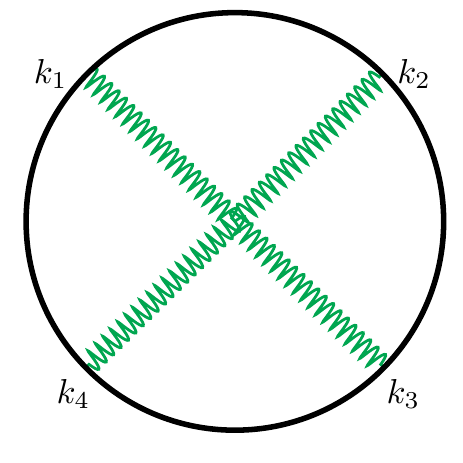}
	\caption{The four point contact diagram.\label{4ptv1}}
\end{figure}

\subsubsection{Final result}
From \equref{\ref{eq: 4s contribution}} and \equref{\ref{eq: 4c contribution}} we can write down the full four point amplitude as
\begin{equation}
\begin{aligned}
\mathcal{M}_{\text{4-pt}}=&\mathcal{M}_s+ \mathcal{M}_t+ \mathcal{M}_{c}\\
=&-\frac{i}{k_{1234}}\bigg[\frac{ V^{12m}(\vect{k}_{1}, \vect{k}_{2}, -\vect{k}_{{12}} )  V^{34n} (\vect{k}_{3}, \vect{k}_{4}, \vect{k}_{{12}} ) }{ k_{12\underline{12}} k_{34\underline{12}}}\left(\eta_{mn}+\frac{k_{1234 \underline{12}}(\vect{k}_{12})_m (\vect{k}_{12})_n}{k_{12} k_{34}  k_{\underline{12}}}\right)\\&\qquad\quad+ \frac{ V^{23m}(\vect{k}_{2}, \vect{k}_{3}, -\vect{k}_{{23}})  V^{41n} (\vect{k}_{4}, \vect{k}_{1}, \vect{k}_{{23}} ) }{ k_{23\underline{23}} k_{41\underline{23}}}\left(\eta_{mn}+\frac{k_{2341 \underline{23}}(\vect{k}_{23})_m (\vect{k}_{23})_n}{k_{23} k_{41}  k_{\underline{23}}}\right)\\&\qquad\quad+V_c^{1234}\bigg]\;.
\end{aligned}\label{eq: Four point amplitude total}
\end{equation}
Here the second line is the $t$ channel contribution which simply reads as
\begin{equation}
\mathcal{M}_{t} = \mathcal{M}_{s} ( 1 \mapsto 2 , 2 \mapsto 3, 3 \mapsto 4, 4 \mapsto 1)\;.
\end{equation}

Note that this result in generic dimensions is noted in eqn.~(6.26) of \cite{Raju:2011mp}, albeit in an integral form\footnote{The author carries out the computation using BCFW-like recursion relation and provides an explicit result in helicity basis in section 5.1.1 of \cite{Raju:2012zs}.}. For completeness, we have carried out those integrations to provide an explicit compact expression for the four point amplitude. Like all other symbolic integrations we have carried out in this paper, we numerically verified these calculations as well.

\subsection{Five point function}
\label{sec:fivepoint}	
\begin{figure}[tbp]
	\centering
	\includegraphics[width=.3\textwidth,origin=c]{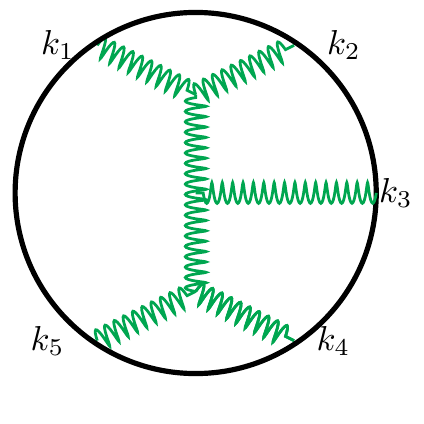}
	\caption{Five point type-a vector diagram.\label{5ptv1}}
\end{figure}

\begin{figure}[tbp]
	\centering
	\includegraphics[width=.3\textwidth,origin=c]{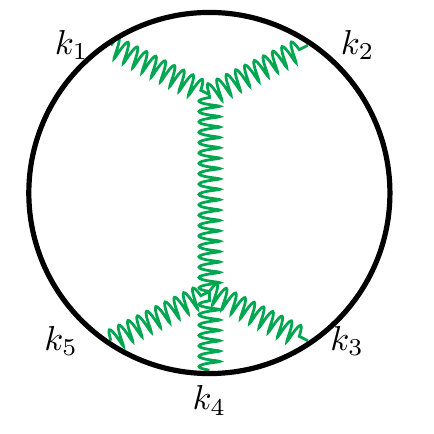}
	\caption{Five point type-b vector diagram.\label{5ptv2}}
\end{figure}

In this section we will write the five-point amplitude of a gauge boson propagating in Anti- de Sitter space at tree level. As in the case of four point, we will look at color-ordered expressions. There are two types of diagrams for the five point amplitude, as depicted in figure~\ref{5ptv1} and figure~\ref{5ptv2}. After permuting these two topologies we can obtain diagrams which can be added to compute the total five point amplitude:
\begin{equation}
\mathcal{M}_{5}=\mathcal{M}_{5a}+\mathcal{M}_{5b}+\text{permutations}
\end{equation}

\subsubsection{Type-a five point diagram}
The amplitude is given as
\begin{multline}
\label{5pt:1}
\mathcal{M}_{5a} =\int (pdp)(qdq)\frac{d z_1}{z_1^{4}}\frac{d z_2}{z_2^{4}}\frac{d z_3}{z_3^{4}} {M}^{12345}_a\left(\vect{k}_1, \vect{k}_2, \vect{k}_3, \vect{k}_4, \vect{k}_5\right)\\
\times\frac{\kkj{k_1,k_2,p,z_1}\kjj{k_3,p,q,z_2}\kkj{k_4,k_5,q,z_3}}{\left( \vect{k}_{12}^2 + p^2\right) \left( \vect{k}_{45}^2 + q^2\right)}
\end{multline}
where
\begin{multline}
{M}^{ijktu}_a(\vect{k}_1, \vect{k}_2, \vect{k}_3, \vect{k}_4, \vect{k}_5 )= V^{ijm}(\vect{k}_1, \vect{k}_2, -\vect{k}_{12} )  H_{mn}(p,\vect{k}_{12})    \\  \times V^{kln} (\vect{k}_3, \vect{k}_{45}, \vect{k}_{12} ) H_{ls}(q,\vect{k}_{45}) V^{stu} (-\vect{k}_{45}, \vect{k}_4, \vect{k}_5 )\;.
\end{multline}

Just like the four point case, we can rewrite it in terms of different pieces of different internal momenta dependence:
\begin{multline}
\mathcal{M}_{5a}=-
V^{12i}(\vect{k}_1, \vect{k}_2, -\vect{k}_{12} ) V^{3kj} (\vect{k}_3, \vect{k}_{45}, \vect{k}_{12} ) V^{l45} (-\vect{k}_{45}, \vect{k}_4, \vect{k}_5 )
\bigg(\eta_{ij}\eta_{kl}\mathcal{M}_{5a}^{(1)}\\+(\vect{k}_{12})_i(\vect{k}_{12})_j \eta_{kl}\mathcal{M}_{5a}^{(2)}+\eta_{ij}(\vect{k}_{45})_k(\vect{k}_{45})_l \mathcal{M}_{5a}^{(3)}+(\vect{k}_{12})_i(\vect{k}_{12})_j(\vect{k}_{45})_k(\vect{k}_{45})_l \mathcal{M}_{5a}^{(4)}\bigg)\;.
\end{multline}

Let us focus on $\mathcal{M}_{5a}^{(1)}$. It reads as
\begin{multline}
\mathcal{M}_{5a}^{(1)}=\int (pdp)(qdq) \left( \frac{\sqrt{\frac{2 p }{\pi }}}{{(k_1+k_2)}^2+p^2} \right) 
\left(\frac{4 k_3 \sqrt{p} \sqrt{q}}{\pi  \left(k_3^2+(p-q)^2\right) \left(k_3^2+(p+q)^2\right)}\right)\\\times
\left( \frac{\sqrt{\frac{2 q }{\pi }}}{k_{45}^2+q^2}  \right) \frac{ 1}{ \vect{k}_{{12}}^2 + p^2 }\frac{ 1}{ \vect{k}_{45}^2 + q^2 }
\end{multline}

One can now proceed and do the brute force integration. Instead, we will find the residues as we did for the four point case and use the residue theorem to get the answer. We find that

\scriptsize
\begin{equation}
\begin{aligned}
\operatorname*{Res}_{\substack{
		p=-i k_{\underline{12}}
		\\
		q=i k_{3 \underline{12}}
}} \mathcal{M}_{5a}^{(1)} &=
\frac{k_{3 \underline{12}}}{2 \pi ^2
	\left(k_{12}^2-k_{\underline{12}}^2\right) \left(k_{3 \underline{12}}^2-k_{45}^2\right) \left(k_{3
		\underline{12}}^2-k_{\underline{45}}^2\right)}
\\
\operatorname*{Res}_{\substack{
		p=-i k_{12}		
		\\
		q=i k_{123}		
}} \mathcal{M}_{5a}^{(1)} &=
-\frac{k_{123}}{2 \pi ^2 \left(k_{123}^2-k_{45}^2\right)
	\left(k_{12}^2-k_{\underline{12}}^2\right) \left(k_{123}^2-k_{\underline{45}}^2\right)}
\\
\operatorname*{Res}_{\substack{
		p=i \left(k_3-k_{\underline{45}}\right)		
		\\
		q=i k_{\underline{45}}		
}} \mathcal{M}_{5a}^{(1)} &=
\frac{k_3-k_{\underline{45}}}{2 \pi ^2
	\left(k_{\underline{45}}^2-k_{45}^2\right) \left(\left(k_{\underline{45}}-k_3\right){}^2-k_{12}^2\right)
	\left(\left(k_{\underline{45}}-k_3\right){}^2-k_{\underline{12}}^2\right)} 
\\
\operatorname*{Res}_{\substack{
		p=i \left(k_3-k_{45}\right)		
		\\
		q=i k_{45}		
}} \mathcal{M}_{5a}^{(1)} &=
\frac{k_3-k_{45}}{2 \pi ^2
	\left(\left(k_{45}-k_3\right){}^2-k_{12}^2\right) \left(\left(k_{45}-k_3\right){}^2-k_{\underline{12}}^2\right)
	\left(k_{45}^2-k_{\underline{45}}^2\right)}
\\
\operatorname*{Res}_{\substack{
		p=i k_{\underline{12}}		
		\\
		q=i k_{3 \underline{12}}		
}} \mathcal{M}_{5a}^{(1)} &=
\frac{k_{3 \underline{12}}}{2 \pi ^2
	\left(k_{12}^2-k_{\underline{12}}^2\right) \left(k_{3 \underline{12}}^2-k_{45}^2\right) \left(k_{3
		\underline{12}}^2-k_{\underline{45}}^2\right)}
\\
\operatorname*{Res}_{\substack{
		p=i k_{\underline{12}}		
		\\
		q=i k_{\underline{45}}		
}} \mathcal{M}_{5a}^{(1)} &=
-\frac{2 k_3 k_{\underline{12}} k_{\underline{45}}}{\pi ^2
	\left(k_{\underline{12}}^2-k_{12}^2\right) \left(k_{\underline{12}\;\underline{45}}^2-k_3^2\right)
	\left(\left(k_{\underline{12}}-k_{\underline{45}}\right){}^2-k_3^2\right) \left(k_{\underline{45}}^2-k_{45}^2\right)}
\\
\operatorname*{Res}_{\substack{
		p=i k_{\underline{12}}		
		\\
		q=i k_{45}		
}} \mathcal{M}_{5a}^{(1)} &=
-\frac{2 k_3 k_{45} k_{\underline{12}}}{\pi ^2
	\left(k_{\underline{12}}^2-k_{12}^2\right) \left(\left(k_{\underline{12}}-k_{45}\right){}^2-k_3^2\right) \left(k_{45
		\underline{12}}^2-k_3^2\right) \left(k_{45}^2-k_{\underline{45}}^2\right)}
\\
\operatorname*{Res}_{\substack{
		p=i k_{12}
		\\
		q=i k_{123}				
}} \mathcal{M}_{5a}^{(1)} &=
-\frac{k_{123}}{2 \pi ^2 \left(k_{123}^2-k_{45}^2\right)
	\left(k_{12}^2-k_{\underline{12}}^2\right) \left(k_{123}^2-k_{\underline{45}}^2\right)}
\\
\operatorname*{Res}_{\substack{
		p=i k_{12}
		\\
		q=i k_{\underline{45}}				
}} \mathcal{M}_{5a}^{(1)} &=
-\frac{2 k_3 k_{12} k_{\underline{45}}}{\pi ^2
	\left(k_{12}^2-k_{\underline{12}}^2\right) \left(\left(k_{12}-k_{\underline{45}}\right){}^2-k_3^2\right)
	\left(k_{\underline{45}}^2-k_{45}^2\right) \left(k_{12 \underline{45}}^2-k_3^2\right)}
\\
\operatorname*{Res}_{\substack{
		p=i k_{12}
		\\
		q=i k_{45}				
}} \mathcal{M}_{5a}^{(1)} &=
-\frac{2 k_3 k_{12} k_{45}}{\pi ^2 \left(\left(k_{12}-k_{45}\right){}^2-k_3^2\right)
	\left(k_{1245}^2-k_3^2\right) \left(k_{12}^2-k_{\underline{12}}^2\right) \left(k_{45}^2-k_{\underline{45}}^2\right)}
\\
\operatorname*{Res}_{\substack{
		p=i k_{3 \underline{45}}
		\\
		q=i k_{\underline{45}}				
}} \mathcal{M}_{5a}^{(1)} &=
\frac{-k_3-k_{\underline{45}}}{2 \pi ^2
	\left(k_{\underline{45}}^2-k_{45}^2\right) \left(k_{3\ \underline{45}}^2-k_{12}^2\right) \left(k_{3
		\underline{45}}^2-k_{\underline{12}}^2\right)} 
\\
\operatorname*{Res}_{\substack{
		p=i k_{345}
		\\
		q=i k_{45}			
}} \mathcal{M}_{5a}^{(1)} &=
-\frac{k_{345}}{2 \pi ^2 \left(k_{345}^2-k_{12}^2\right)
	\left(k_{345}^2-k_{\underline{12}}^2\right) \left(k_{45}^2-k_{\underline{45}}^2\right)}
\end{aligned}\label{explicit residues of five point function}
\end{equation}		
\normalsize

We can now sum these residues and write the total result:
\begin{equation}
\mathcal{M}_{5a}^{(1)}=(i\pi)^2\left(\sum\limits_i\operatorname*{Res}_{p_i,q_i}\right)\mathcal{M}_{5a}^{(1)}
=
\frac{k_{12 \underline{12}3345\underline{45}}}{k_{12345} k_{12 \underline{12}} k_{345 \underline{12}} k_{3 \underline{12}\;\underline{45}} k_{45
		\underline{45}} k_{123 \underline{45}}}
\end{equation}
This answer is remarkably simpler than the  individual residues.

We can similarly do the other integrations:
\begin{equation}
\begin{aligned}
\mathcal{M}_{5a}^{(2)}
&=\frac{k_1 k_{3345 \underline{12.45}} k_{1223345 \underline{12.45}}+k_{2345} k_{23 \underline{45}} k_{3345\
		\underline{45}}+k_{\underline{12}} k_{23345 \underline{12.45}} k_{23345 \underline{45}}}{k_{12} k_{345} k_{12345}
	k_{\underline{12}} k_{12\ \underline{12}} k_{345\ \underline{12}} k_{3\ \underline{45}} k_{45 \underline{45}} k_{123\
		\underline{45}} k_{3 \underline{12}\;\underline{45}}}
\\
\mathcal{M}_{5a}^{(3)}
&=\frac{k_{5312} k_{53 \underline{12}} k_{3312 \underline{12}}+k_4 k_{3312 \underline{12}\;\underline{45}} k_{1223345
		\underline{12.45}}+k_{53312 \underline{12}} k_{53312\underline{12}\;\underline{45}} k_{\underline{45}}}{k_{45} k_{312} k_{12345}
	k_{3 \underline{12}} k_{12 \underline{12}} k_{453 \underline{12}} k_{\underline{45}} k_{45 \underline{45}} k_{312
		\underline{45}} k_{3\underline{12}\;\underline{45}}}
\\
\mathcal{M}_{5a}^{(4)}
&= \frac{1}{k_{12} k_{45} k_{123} k_{345} k_{12 \underline{12}} k_{345
		\underline{12}} k_{45 \underline{45}} k_{123 \underline{45}}}\left[\frac{1}{k_{12345} }+\widehat{\mathcal{M}}_{5a}^{(4)}\right]
\end{aligned}
\end{equation}
where $\widehat{\mathcal{M}}_{5a}^{(4)}$ is a genuinely AdS term that does not contribute in flat space limit. Its explicit form is as follows.
\begin{equation}
\scriptsize
\begin{aligned}
\widehat{\mathcal{M}}_{5a}^{(4)}=&\left(k_2+2 k_3+k_{\underline{45}}\right) \left(k_3+k_4+k_5+k_{\underline{45}}\right)
k_{\underline{12}}^3+\bigg[6 k_3^3+4 \left(2 k_4+2 k_5+3 k_{\underline{45}}\right)
k_3^2\\&+\left(2 k_4^2+4 k_5 k_4+8 k_{\underline{45}} k_4+2 k_5^2+7 k_{\underline{45}}^2+8
k_5 k_{\underline{45}}\right) k_3+k_{\underline{45}}
\left(k_4+k_5+k_{\underline{45}}\right){}^2+k_2^2
\left(k_3+k_4+k_5+k_{\underline{45}}\right)\\&+k_2 \left(5 k_3^2+\left(6 k_4+6 k_5+7
k_{\underline{45}}\right) k_3+k_4^2+k_5^2+2 k_{\underline{45}}^2+2 k_4 k_5+3 k_4
k_{\underline{45}}+3 k_5 k_{\underline{45}}\right)\bigg]
k_{\underline{12}}^2\\&+\bigg[\left(3 k_3^2+4 \left(k_4+k_5+k_{\underline{45}}\right)
k_3+\left(k_4+k_5+k_{\underline{45}}\right){}^2\right) k_2^2+\bigg\{8 k_3^3+3 \left(4
k_4+4 k_5+5 k_{\underline{45}}\right) k_3^2\\&+4 \left(k_4^2+2 k_5 k_4+3 k_{\underline{45}}
k_4+k_5^2+2 k_{\underline{45}}^2+3 k_5 k_{\underline{45}}\right) k_3+k_{\underline{45}}
\left(2 k_4^2+4 k_5 k_4+3 k_{\underline{45}} k_4+2 k_5^2+k_{\underline{45}}^2+3 k_5
k_{\underline{45}}\right)\bigg\} k_2\\&+\left(2 k_3+k_{\underline{45}}\right) \bigg\{3
k_3^3+\left(5 k_4+5 k_5+6 k_{\underline{45}}\right) k_3^2+\left(2 k_4^2+4 k_5 k_4+5
k_{\underline{45}} k_4+2 k_5^2+3 k_{\underline{45}}^2+5 k_5 k_{\underline{45}}\right)
k_3\\&+\left(k_4+k_5\right) k_{\underline{45}}
\left(k_4+k_5+k_{\underline{45}}\right)\bigg\}\bigg]
k_{\underline{12}}+\left(k_2+k_3\right) \left(k_2+k_3+k_{\underline{45}}\right) \bigg\{2
k_3^3+4 \left(k_4+k_5+k_{\underline{45}}\right) k_3^2\\&+2
\left(k_4+k_5+k_{\underline{45}}\right){}^2 k_3+\left(k_4+k_5\right) k_{\underline{45}}
\left(k_4+k_5+k_{\underline{45}}\right)\bigg\}+k_1^2 \bigg[2 k_3^3+4
\left(k_4+k_5+k_{\underline{45}}\right) k_3^2\\&+2
\left(k_4+k_5+k_{\underline{45}}\right){}^2 k_3+\left(k_4+k_5\right) k_{\underline{45}}
\left(k_4+k_5+k_{\underline{45}}\right)+k_{\underline{12}}^2
\left(k_3+k_4+k_5+k_{\underline{45}}\right)\\&+k_{\underline{12}} \left(3 k_3^2+4
\left(k_4+k_5+k_{\underline{45}}\right)
k_3+\left(k_4+k_5+k_{\underline{45}}\right){}^2\right)\bigg]+k_1 \left(2 k_2+2
k_3+k_{\underline{12}}+k_{\underline{45}}\right) \bigg[2 k_3^3\\&+4
\left(k_4+k_5+k_{\underline{45}}\right) k_3^2+2
\left(k_4+k_5+k_{\underline{45}}\right){}^2 k_3+\left(k_4+k_5\right) k_{\underline{45}}
\left(k_4+k_5+k_{\underline{45}}\right)+k_{\underline{12}}^2
\left(k_3+k_4+k_5+k_{\underline{45}}\right)\\&+k_{\underline{12}} \left(3 k_3^2+4
\left(k_4+k_5+k_{\underline{45}}\right)
k_3+\left(k_4+k_5+k_{\underline{45}}\right){}^2\right)\bigg]  \;.
\end{aligned}\normalsize
\end{equation}			

\subsubsection{Type-b five point diagram}
The amplitude is given as
\begin{multline}
\label{5pt:2}
\mathcal{M}_{5b} =\int (pdp)\frac{d z_1}{z_1^{4}}\frac{d z_2}{z_2^{4}} {M}_b^{12345}\left(\vect{k}_1, \vect{k}_2, \vect{k}_3, \vect{k}_4, \vect{k}_5\right)\frac{\kkkj{k_1,k_2,k_3,p,z_1}\kkj{k_4,k_5,p,z_2}}{ \left( \vect{k}_{45}^2 + p^2\right)}
\end{multline}
for
\begin{equation}
{M}^{ijklm}_b(\vect{k}_1, \vect{k}_2, \vect{k}_3, \vect{k}_4, \vect{k}_5 )= V^{ija}(\vect{k}_1, \vect{k}_2, -\vect{k}_{12} )
H_{ab}(p,\vect{k}_{12}) V^{klmb}\;,
\end{equation}
where $V^{klmb}$ was defined in \equref{\ref{eq: v structures}} .

The integrations over $z_1$ and $z_2$ can immediately be carried out via \equref{\ref{integratedz}}. Then one can do the $p-$integration, either directly or via the method of residues. The result takes the form
\begin{equation}
\mathcal{M}_{5b} =-i\frac{V^{12a}(\vect{k}_{1}, \vect{k}_2, -\vect{k}_{12} ) V^{345b} }{k_{12345} k_{45\underline{45}} k_{123 \underline{45}}}\left(\eta_{ab}+\frac{k_{12345\underline{45}}}{k_{123}k_{\underline{45}}k_{45}}(\vect{k}_{45})_a(\vect{k}_{45})_b\right)\;.
\end{equation}	

We can now write the full five point amplitude as the summation over different topologies and their permutations. We can compare this with the case of four point amplitude in \equref{\ref{eq: Four point amplitude total}}, where we have two topologies and two permutations for the first one (first two lines there) and one permutation for the second one (last line).

\subsection{Six point function}
\label{sec:sixpoint}
Let us move on to six point amplitudes. Unlike the previous cases, we will not be providing the explicit results here because of two reasons: Firstly, the number of calculations proliferates as we consider higher point amplitudes; in particular, we have to compute 18 different terms to fully calculate all the topologies. While doing the calculations is actually straightforward and can easily be automated, the results are not extremely illuminating and involve lengthy pieces such as $\widehat{\mathcal{M}}_{5a}^{(4)}$.

The second reason for us to avoid explicitly writing six point amplitude is that we expect a pattern and we believe higher point calculations can be carried out more efficiently for generic $n$ point amplitude.\footnote{Here, we mean symbolic calculation as our method (or even a direct integration) is already sufficiently convenient for numerical studies.} We do not pursue this idea in present work; instead, we give a taste of what the six point amplitude looks like by computing one of the terms appearing in type-a topology that is to be defined below: Other expressions can be calculated in a similar fashion.

\begin{figure}[tbp]
	\centering
	\includegraphics[width=.8\textwidth,origin=c]{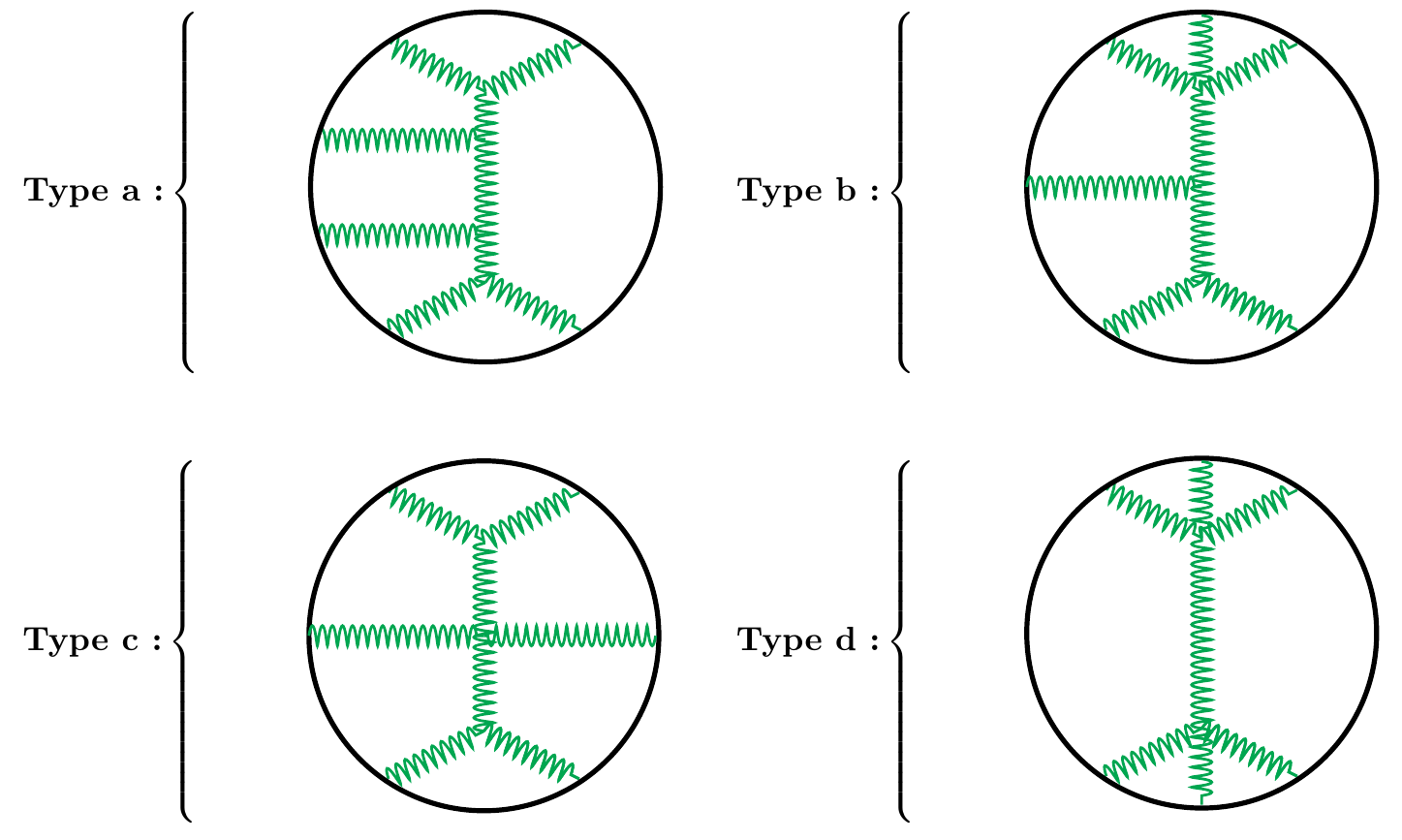}
	\caption{Topologies for six point vector diagrams. \label{6pt}}
\end{figure}

In six points, we have 4 different topologies:
\begin{equation}
\mathcal{M}_{6}=\mathcal{M}_{6a}+\mathcal{M}_{6b}+\mathcal{M}_{6c}+\mathcal{M}_{6d}+\text{permutations}
\end{equation}
where we can see them as in figure~\ref{6pt}. In particular, we can write type-$a$ amplitude as
\begin{multline}
\label{6pt:1}
\mathcal{M}_{6a} =\int (p_1dp_1)(p_2dp_2)(p_3dp_3)\frac{d z_1}{z_1^{4}}\frac{d z_2}{z_2^{4}}\frac{d z_3}{z_3^{4}}\frac{d z_4}{z_4^{4}} {M}^{123456}_a\left(\vect{k}_1, \vect{k}_2, \vect{k}_3, \vect{k}_4, \vect{k}_5, \vect{k}_6\right)\\
\times\frac{\kkj{k_1,k_2,p_1,z_1}\kjj{k_3,p_1,p_2,z_2}\kjj{k_4,p_2,p_3,z_4}\kkj{k_5,k_6,p_3,z_4}}{\left( \vect{k}_{12}^2 + p_1^2\right) \left( \vect{k}_{123}^2 + p_2^2\right)\left( \vect{k}_{56}^2 + p_3^2\right)}
\end{multline}
where ${M}^{ijklmn}_a$ is of the form
\begin{multline}
{M}^{ijklmn}_a(\vect{k}_1, \vect{k}_2, \vect{k}_3, \vect{k}_4, \vect{k}_5, \vect{k}_6 )=V^{ija}(\vect{k}_1, \vect{k}_2,-\vect{k}_{12})H_{ab}(p_1,\vect{k}_{12})V^{kbc}(\vect{k}_3, \vect{k}_{12},-\vect{k}_{123})\\\times H_{cd}(p_2,\vect{k}_{123})V^{lde}(\vect{k}_4, \vect{k}_{123},\vect{k}_{56})H_{ef}(p_3,\vect{k}_{56})V^{mnf}(\vect{k}_5, \vect{k}_{6},-\vect{k}_{56})
\end{multline}
hence
\begin{multline}
\mathcal{M}_{6a} =-i\;V^{12a}(\vect{k}_1, \vect{k}_2,-\vect{k}_{12})\eta_{ab}V^{3bc}(\vect{k}_3, \vect{k}_{12},-\vect{k}_{123})\eta_{cd} \\\times V^{4de}(\vect{k}_4, \vect{k}_{123},\vect{k}_{56})\eta_{ef} V^{56f}(\vect{k}_5, \vect{k}_{6},-\vect{k}_{56}) M_{6a}^{(1)}+\cdots
\end{multline}

Since the result is relatively simple, we will be providing $M_{6a}^{(1)}$ here:
\begin{equation}
M_{6a}^{(1)}=\frac{1}{k_{12 \underline{12}} k_{3456 \underline{12}} k_{56 \underline{56}} k_{1234
		\underline{56}} k_{123 \underline{123}} k_{456 \underline{123}}}\bigg(\frac{1}{k_{123456}}+\frac{k_{\underline{12} 12\underline{123}334\underline{56} } k_{ \underline{12}3\underline{123}4456
		\underline{56} }}{ k_{ \underline{12}34 \underline{56}} k_{\underline{12} 3\underline{123}} k_{\underline{123}4\underline{56}}}\bigg)\;.
\end{equation}		

\section{Conclusion}
\label{sec:conclusion}
In this paper, we have calculated tree level momentum space amplitudes for a massless vector propagating in AdS$_4$. We are interested in such correlators because of their physical relevance and their simplicity. 
These momentum correlators have not been fully investigated despite their relevance in cosmology, conformal theories, and their possible connection with scattering amplitudes in flat space. 

We have computed explicit examples of previously uncomputed higher point correlators in Anti-de Sitter space.  We found it convenient to work in the axial gauge as it allowed us to set the radial component to zero. We explicitly computed  four and  five point amplitudes and provided a recipe to compute full six point expression. Remarkably, we discovered that despite the tediousness of computation, we were able to write a compact and relatively simple expression.

We believe that with this paper, we built on the formalism developed by Raju to provide a complementary calculation to \cite{Raju:2012zs,Raju:2012zr}, which were done using BCFW-like all line recursion relations. While these recursion relations can be used in practice to compute higher point amplitudes, the number of required partitions for the computation is greater than the usual BCFW. The reason for this difference is the way external momenta are deformed: Unlike BCFW in which only two of them are complexified, all external lines are deformed in \cite{Raju:2012zr}, similar to Risager recursion relations introduced in \cite{Risager:2005vk}. This difference also leads to quadratic solutions for the poles, meaning that the answers become extremely complicated at five point and higher. This motivated us to compute higher point correlation functions directly.

We would like to present many promising future directions. As the expressions for these correlators are simple, we believe that there must be a simple structure that can recursively compute $n$-point correlation function. We believe that these structures have similar properties as the ones calculated in \cite{Arkani-Hamed:2017fdk}. Work in this direction is in progress {\cite{ChandraSavan}} where we also hope to understand the flat space limits better. It would also be interesting to consider other spacetime dimensions or higher spin correlators in AdS$_4$. 

\acknowledgments
We thank David Meltzer, Anindya Dey, and Chandramouli Chowdhury for helpful discussions. SA is supported by NSF grant PHY-1350180 and Simons Foundation grant 488651.

\bibliography{savanreference}{}
\bibliographystyle{JHEP}
\end{document}